\documentclass[letterpaper]{mn2e}

\usepackage{amssymb}
\usepackage{graphicx}
\usepackage{psfrag}

\def\Vsigma{\ifmmode{V/\sigma}\else{$V/\sigma$}\fi}
\def\h3{\ifmmode{h_{3}}\else{$h_{3}$}\fi}
\def\Reff{\ifmmode{R_\mathrm{eff}}\else{$R_\mathrm{eff}$}\fi}

\begin{document}

\input epsf

\title[Line-of-sight velocity distributions of elliptical galaxies from mergers]
      {Line-of-sight velocity distributions of elliptical galaxies from collisionless mergers}

\author[A. C. Gonz\'alez-Garc\'{\i}a, M. Balcells \& V.~S. Olshevsky]{A. C. Gonz\'alez-Garc\'{\i}a$^1$, M. Balcells$^1$ \& V.~S. Olshevsky$^{1,2}$
\\$ ^1$ Instituto de Astrof\'{\i}sica de Canarias, C/ V\'{\i}a Lactea s/n, 38200 La Laguna, Tenerife, Spain
\\$ ^2$ Main Astronomical Observatory, 27 Akademika Zabolotnoho St., 03680, Kiev, Ukraine}

\maketitle

\begin{abstract}

We analyse the skewness of the line-of-sight velocity distributions in model elliptical galaxies built through collisionless galaxy mergers.  We build the models using large N-body simulations of mergers between either two spiral or two elliptical galaxies.  
Our aim is to investigate whether the observed ranges of  skewness coefficient (\h3)  and the rotational support (\Vsigma), as well as the anticorrelation between \h3\ and $V$, may be reproduced through collisionless mergers.
Previous attempts using N-body simulations failed to reach $\Vsigma \approx 1-2$ and corresponding  high \h3\ values, which suggested that gas dynamics and ensuing star formation  might be needed in order to explain the skewness properties of ellipticals through mergers.  
Here we show that high \Vsigma\ and high \h3\ are reproduced in collisionless spiral-spiral mergers whenever a central bulge allows the discs to retain some of their original angular momentum during the merger.   We also show that elliptical-elliptical mergers, unless merging from a high-angular momentum orbit,  reproduce the strong skewness observed in non-rotating, giant, boxy ellipticals.   The behaviour of the \h3\ coefficient therefore associates rapidly-rotating disky ellipticals to disc-disc mergers, and associates boxy, slowly-rotating giant ellipticals to elliptical-elliptical mergers, a framework generally consistent with the expectations of hierarchical galaxy formation.  

\end{abstract}


\section{Introduction}
\label{sec:introduction}

Elliptical galaxies are frequently divided into two sets with different characteristics. High luminosity ellipticals showing boxy isophotes are mainly supported by velocity dispersion, and they are commonly associated to extended X-ray haloes and are radio loud. Low luminosity elliptical galaxies present a high amount of rotation being rotationally supported. They show disky deviations in their elliptical isophotes and usually they are not associated to X-ray emission nor to radio loud objects (Lauer 1985, Bender 1988, Bender et al. 1989).

Explaining such dichotomy is a basic requirement of elliptical galaxy formation theories.  Early N-body work, which pointed out that 1:1 collisionless disc galaxy mergers give as a result a body that resembles a boxy, slow-rotator elliptical galaxy, while mergers with mass ratios of the order 3:1 give disky ellipticals (Barnes 1998; Naab et al. 1999; Bendo \& Barnes 2000, hereafter BB00; Naab \& Burkert 2001, 2003) ignored the basic fact that the two classes of ellipticals are mainly segregated by luminosity, and hence, mass.   In hierarchical scenarios for galaxy growth, because successive merging moves galaxies into progressively bulge-dominated systems, elliptical galaxies are end-points, hence they dominate the high end of the mass distribution.  In these schemes, massive $z=0$ ellipticals, would be expected to form through mergers of ellipticals.   Because of this, explanations for the dichotomy in elliptical galaxy properties in a hierarchical context will naturally be related to differences between the galaxy types of the precursors.  

\begin{figure}
\centering
\scalebox{0.4}{\includegraphics{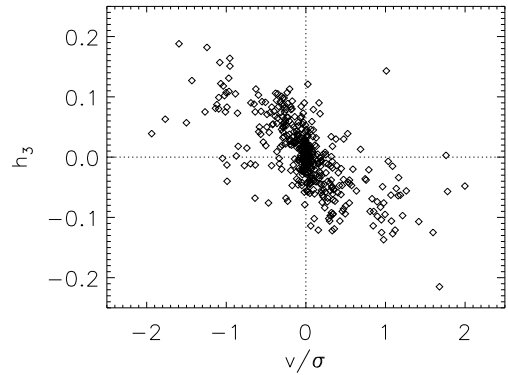}}
\caption{$h_3$ vs. $V/\sigma$ for 44 near-by elliptical galaxies from BSG. \label{f0}}
\end{figure}

In this letter, we elaborate on the types of precursors that may have merged to form elliptical galaxies by analysing  the line-of-sight velocity distributions (LOSVD) in simulated N-body merger remnants.  LOSVDs of elliptical galaxies may be obtained from high signal-to-noise (S/N) spectroscopy, and are customarily modeled by fitting a Gauss-Hermite series (e.g., van der Marel \& Franx 1993).  Two indices, \h3\ and $h_{4}$, parametrize the strength of  asymmetric and symmetric deviations from Gaussian shape, respectively.  
Bender, Saglia \& Gerhard (1994, hereafter BSG) show that \h3, the skewness of the LOSVD, has opposite sign to the rotation velocity: the LOSVD is steep in the leading side of the rotation, and has a more extended tail on the trailing side.  Furthermore, BSG show that for ellipticals as a class, \h3\ and the rotational support parameter \Vsigma\  follow a distinct pattern.  We show  their main result (data kindly provided by R. Saglia) in Figure~\ref{f0}, which plots the skewness coefficient \h3\ against \Vsigma, for a number of galaxies; each point corresponds to a local position along a major-axis slit, and all galaxies are shown in the same diagram.  
Clearly, \h3\ and \Vsigma\ have opposite signs.  At low $|\Vsigma|$, which corresponds to giant ellipticals and to the central parts of other ellipticals, $|\h3|$ increases rapidly with $|\Vsigma|$.  At high $|\Vsigma|$, which maps intermediate and outer radii in intermediate-luminosity ellipticals, \h3 reaches up to $|\h3|\sim 0.2$.  
 
Several works have studied what types of LOSVD are generated by mergers.  N-body studies are useful tools as they allow to dissect the remnant LOSVDs into components by, e.g., the contribution by each merger precursor, or by the contribution of each orbit type.  The first study of LOSVDs of N-body merger remnants (Balcells 1991)  showed that mergers of unequal ellipticals lead to asymmetric LOSVDs similar to those observed in kinematically decoupled cores (steep leading sides) and that the asymmetry is not due to an accreted disc but to dynamical transformations in the primary as a result of the merger.   
Bendo \& Barnes (2000, hereafter BB00) pioneered the use of Gauss-Hermite analysis of LOSVDs in N-body merger remnants, and demonstrated that the shape of the LOSVD is intimately linked to the orbital makeup of the remnant.  In short, box orbits have no net rotation, hence their contribution to the LOSVD is fairly symmetric and centered on $V=0$.  Short-axis tubes show net rotation in merger remnants, hence their contribution to the LOSVD peaks at $V > 0$.  Whenever short-axis tube orbits dominate over box orbits, the global LOSVD will be steep on the leading side (leading to LOSVD similar to those observed by BSG), while dominance of box-orbits leads to a peak at $V\sim 0$ and an extended tail on the leading side.

Naab \& Bukert (2001, hereafter NB01) were the first to point out that collisionless major mergers of disc galaxies (mass ratios 1:1 and 3:1; disc:bulge ratio 3:1) lead to LOSVDs with opposite behaviour to observations: \h3\ has the same sign as \Vsigma.   
They noted that their models can be made to agree with the observational trends (opposite signs for \h3\ and \Vsigma) by postulating that ellipticals harbor large-scale stellar discs (15\% in mass, scale-length $h\sim \Reff$) formed after the merger from left-over gas.  The statistics of isophotal diskiness suggest that such discs might in fact exist in most intermediate-luminosity ellipticals (Rix \& White 1990).  

The dissection of LOSVDs into their orbit constituents discussed above  clarifies why collisionless major mergers lead to the wrong correlation between \h3\ and \Vsigma: such merger models have too many box orbits.   The discs postulated by NB01 simply add short-axis tubes and compensate for the excess of box orbits.  It is useful to enquire about ways to increase the relative proportion of z-tube orbits which are more directly related to the merger dynamics.  Box orbits are favored by cores with shallow density profiles, hence the limited spatial resolution of the simulations might lead to too many box orbits.  Limited resolution of the central parts also causes an earlier disruption of the merging galaxies, and to a remnant which is less supported by net rotation, again enhancing box orbits.  
Recent results on collisionless mergers of discs (Gonz\'alez-Garc\'{\i}a \& Balcells 2005, hereafter GGB05) show that the bulge component helps stabilizing the disc during the merger. 
Right before merging, the progenitor discs are still largely in place, delivering more angular momentum to the remnant.  That could enhance the proportion of short-axis tube orbits in the remmnant.  Finally, it should be noted that, in the study of this question, mergers between elliptical galaxies have not been considered so far.

In the present letter we consider merger simulations involving two disc galaxies with large bulges or two elliptical galaxies, varying the initial orbital parameters of the encounters.
In Section~\ref{sec:models} we present our simulations, in Section~\ref{sec:losvds} we explain the LOSVD derivation procedure, Section~\ref{sec:results} presents our results and finally in Section~\ref{sec:disc} we discuss the implications from our results for the current view of elliptical galaxy formation.

\section{Models}\label{sec:models}

The disc galaxy models comprise a halo, a disc and a bulge built following Kuijken \& Dubinski (1995). Masses are in the ratios halo:disc:bulge = 36:5:1 and 19:2:1; other model parameters are those of the $dbh$ models of GGB05, but with a larger number of particles.  We note that the initial mass ratio between disc and bulge is 5:1 for one of the models and 2:1 for the other, this is a somewhat larger bulge than used by Naab \& Burkert (2001, 2003).
Models with different masses are scaled up version of low mass systems following a Tully-Fisher like relation.

The elliptical galaxy models are Jaffe (1983) non-rotating isotropic models without a dark matter halo.  
For details, see Gonz\'alez-Garc\'{\i}a \& van Albada 2005a.  
The Jaffe scale parameter was set to $r_{\rm J} = 1$; truncation at $r= 10 \times r_{\rm J}$ means that the true half-mass radius is $r_{1/2} = 0.82$. 
Systems of masses different from one are scaled up versions of $M=1$ systems following a Fish (1964) law. The initial model was let to evolve in isolation for 10 crossing times. After a mild re-organization, the system reaches equilibrium. This equilibrium configuration is the one used in the merger models.  

Model units are $M_\mathrm{disc} = 1$, $h_\mathrm{disc} = 1$ and $G=1$.  
A set of physical units that match the {\it dbh} models to the Milky Way are:
$[M] = 3.24 \times 10^{11}  \; \rm{M_{\odot}}$; $[L] = 14.0  \rm  \; {kpc}$; $[T] =  4.71\times10^7 \; \rm{yr}$, with $[v] =  315  \rm  \; {km/s}$. For scaling our Jaffe models to elliptical galaxies the set of units would be: $[M] = M_{\rm J} = 4\times10^{11} \; \rm{M_{\odot}}$; $[L] = r_{\rm J} = 10 \;\rm{kpc}$; $[T] =  2.4\times10^7 \;\rm{yr}$, and finally $[v] =  414\; \rm{km/s}$. 

\begin{table}
\begin{center}
\caption{Initial configurations for the models. (1) Model name, (2) type of merger, disc-discs (S+S) or elliptical-elliptical (E+E), (3) Mass ratio, (4) initial orientation of the spin, (5) initial separation, (6) orbit ellipticity and (7)
pericenter distance.  The latter two parameters refer to the Keplerian orbit of
same initial conditions. \label{tab}}

\begin{tabular}{@{}ccccccc}
\hline

{\bf Mod.} &{\bf Type}&{\bf $\frac{M_1}{M_2}$}& {\bf $(\theta_1,\phi_1)$ $(\theta_2,\phi_2)$} & {\bf $r_i$} & {\bf $e$}& {\bf $r_{p}$}\\
(1) & (2) & (3) & (4) & (5) & (6) & (7)\\
\hline
$5d11$ & S+S & 1:1 & (10,-10);(70,30) & 11.25 & 0.7 & 3.38 \\
$5d31$ & S+S & 3:1 & (10,-10);(70,30) & 26.98 & 0.7 & 8.07 \\
$2d11$ & S+S & 1:1 & (10,-10);(70,30) & 11.25 & 0.7 & 3.38 \\
$2d31$ & S+S & 3:1 & (10,-10);(70,30) & 26.98 & 0.7 & 8.07 \\
$e11$& E+E & 1:1 & (0,0);(0,0) & 30 & 1 & 0.08 \\
$e31$& E+E & 3:1 & (0,0);(0,0) & 30 & 1 & 0.15 \\
$e51$& E+E & 5:1 & (0,0);(0,0) & 40 & 1 & 0.05 \\
$e11d$& E+E & 1:1 & (0,0);(0,0) & 30 & 1 & 0.80 \\
\hline

\end{tabular}
\end{center}
\end{table}

Parameters for the mergers analysed here are given in Table~\ref{tab}.  
All experiments involve direct orbits.
We include four spiral-spiral mergers, and four elliptical-elliptical mergers.  For the discs, we chose slightly sub-parabolic orbits to accelerate merging. Remnants of S+S mergers with mass ratios 4:1 to 7:1 are reported to have characteristics between those of ellipticals and spirals (Naab \& Burkert 2003, Bournaud et al. 2004) thus a choice of mass ratios 1:1 to 3:1 seems adequate for this study. For the ellipticals, merger orbits were parabolic. Two equal-mass E+E mergers with different orbital angular momentum were computed, to investigate the dependence of the LOSVDs on the orbital angular momentum.

For the disc models we use $2.75 \times 10^5$ particles for the halo, $1.75 \times 10^5$ particles for the disc and $9 \times 10^4$ for the bulge. This makes a total of $10^6$ particles for the merger remnant. Each elliptical galaxy is modeled with $10^5$ particles. We run the simulations with the parallel version of GADGET (Springel et al. 2001, Springel 2005) on the IAC Beowulf Cluster with 16 CPU's and on the MareNostrum cluster of the Barcelona Supercomputer Center\footnote{\texttt{http://www.bsc.es}}. Energy conservation is good, and variations are kept below the $0.5 \%$. Independent softening is used for each particle type. The softening is taken as one fifth of the half mass radius of the component. After merger is completed we let the remnants to evolve in isolation for 10 half-mass radius crossing times to ensure that the inner parts are close to virial equilibrium.

\section{LOSVD derivation}\label{sec:losvds}

We obtain LOSVDs for the merger remnants in the following way. We choose a point of view at random and project the particle distribution. We then derive a surface density plot by binning the space and counting the number of projected particles in each bin. Finally we define isodensity contours and fit ellipses deriving values for the ellipticity, position angle and the $a_4$ parameter. Then we place a slit along the major axis of our ellipses up to one effective radius \Reff.  We have binned the slit in ten spatial bins and the velocity interval in 50 bins. We find the radial projected velocity and the number of particles in each bin in velocity for each bin in the slit. In this way we obtain a line-of-sight velocity distribution.  

Finally we fit the LOSVD by a Gaussian and the residuals by a Gauss-Hermite polynomial as given by van der Marel \& Franx (1993) and BSG. We repeat this process for each remnant for 90 randomly chosen points of view.  

From the fitting procedure we obtain a value for $V$, the velocity centroid of the distribution at each spatial bin along the slit, the velocity dispersion $\sigma$, and the amplitude of the third Hermite polynomial $h_3$, which is a measurment of the skewness of the distribution.

\begin{figure}
\centering
\scalebox{0.20}{ \includegraphics{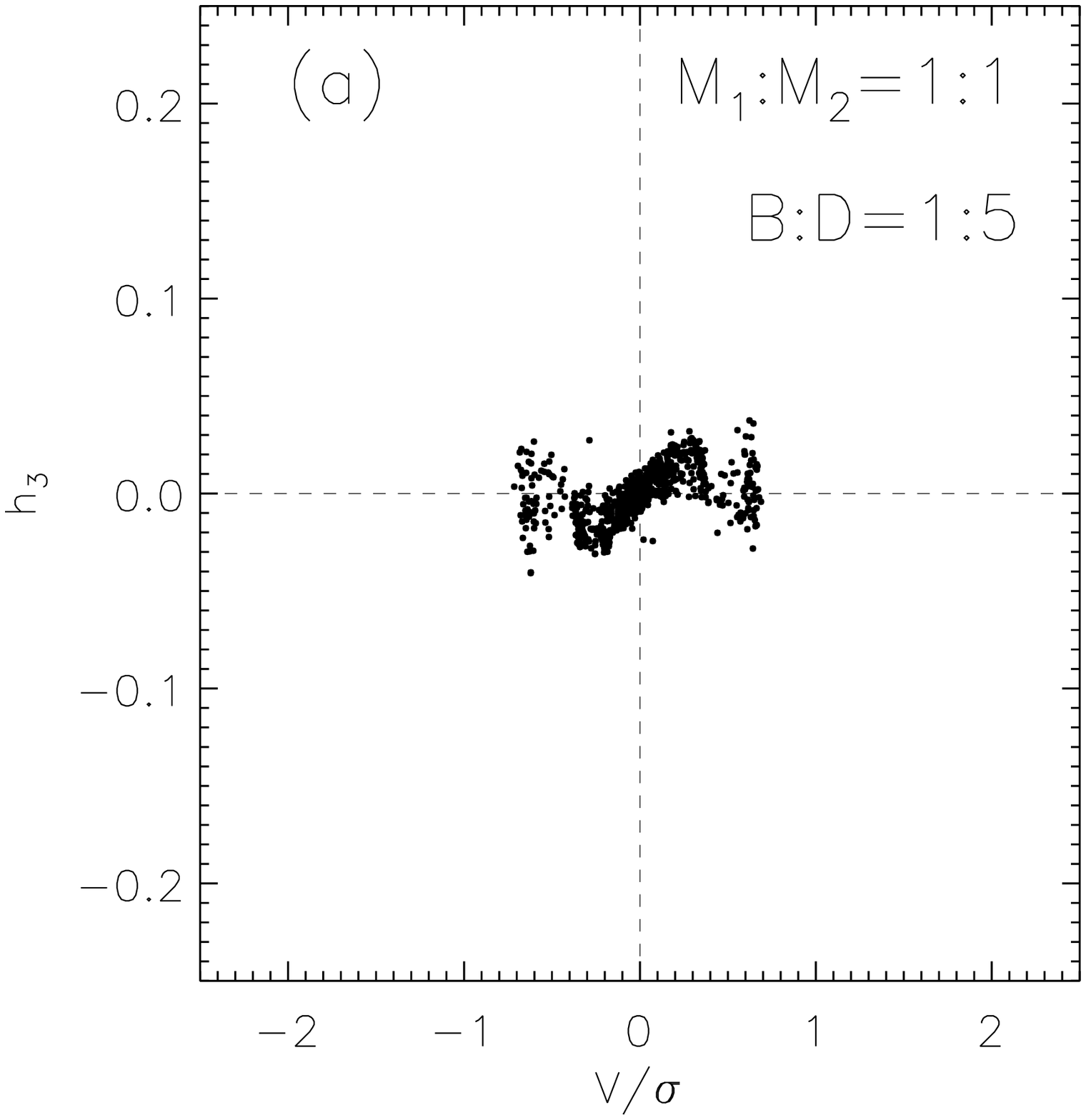}}
\hspace{0cm}
\scalebox{0.20}{ \includegraphics{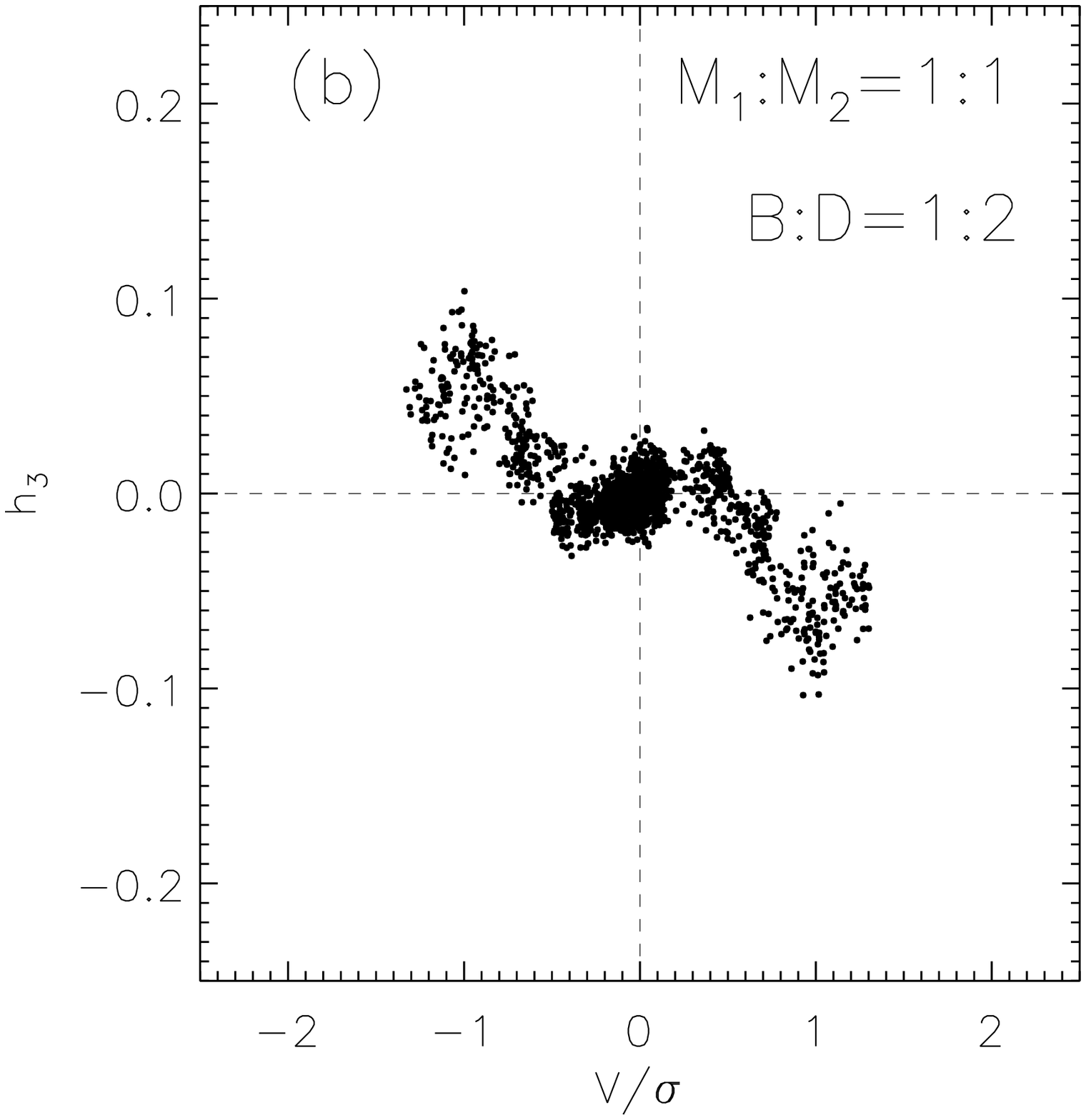}}
\vspace{0cm}
\scalebox{0.20}{ \includegraphics{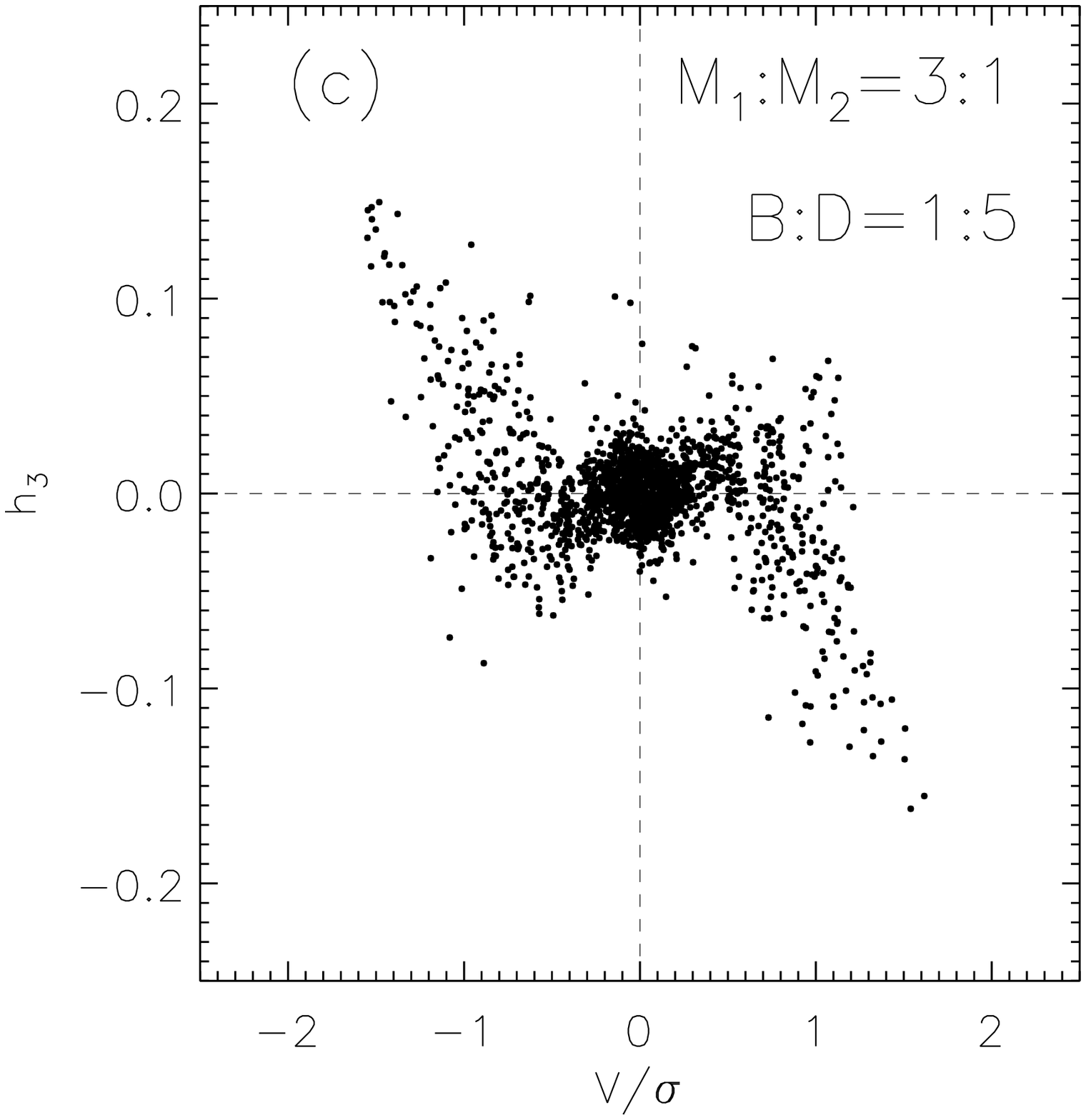}}
\hspace{0cm}
\scalebox{0.20}{ \includegraphics{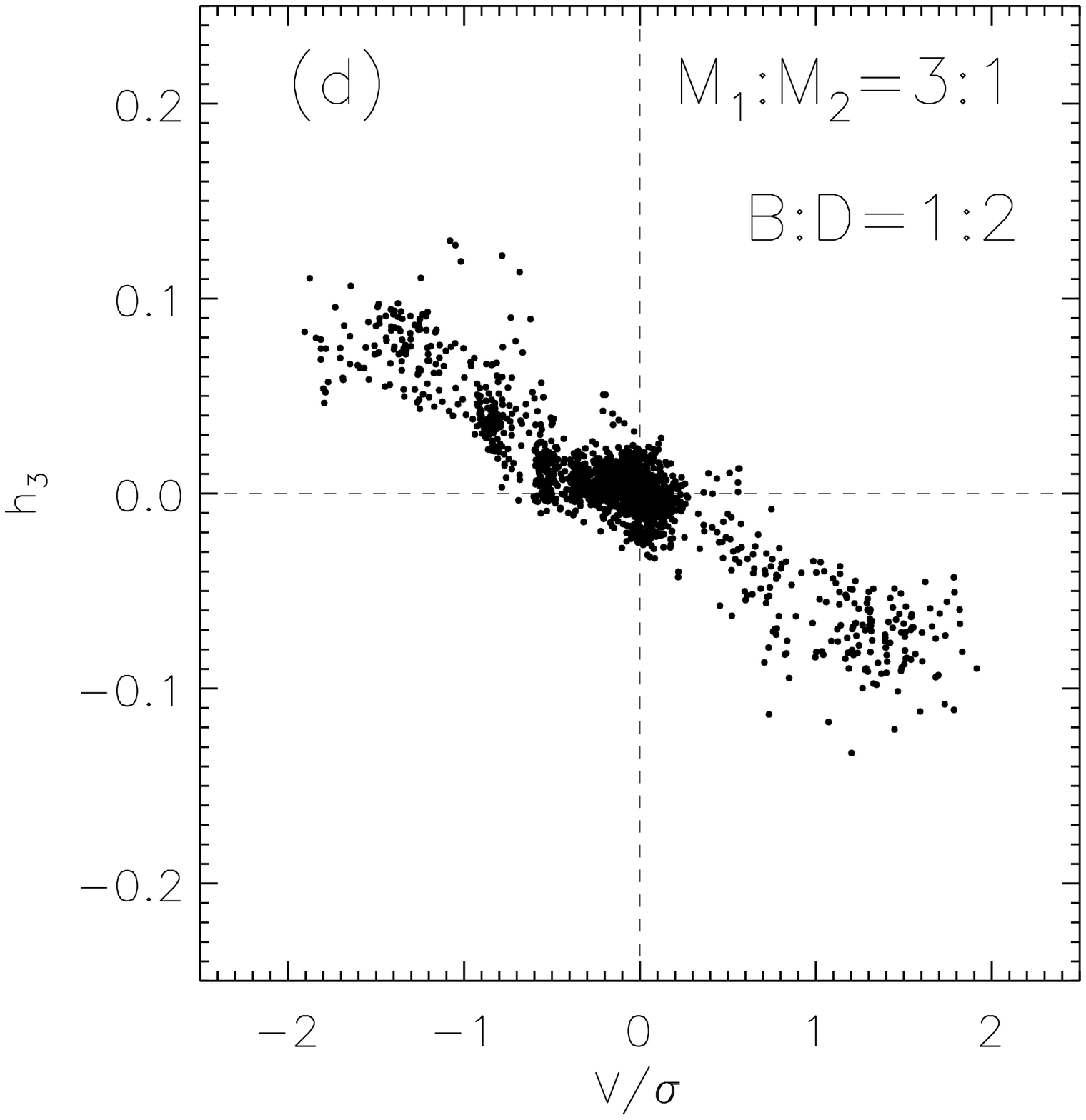}}
\caption{ $h_3$ vs. $V/\sigma$ for the disc mergers. Top row models with  equal mass mergers, bottom row 3:1 merger. Left panels bulge-to-disc ratio 1:5, right models with B:D of 1:2 (a) Model {\it 5d11}. (b) Model {\it 2d11}. (c) Model {\it 5d31}. (d) Model {\it 2d31}.\label{f1}}
\end{figure}

\begin{figure}
\centering
\scalebox{0.20}{ \includegraphics{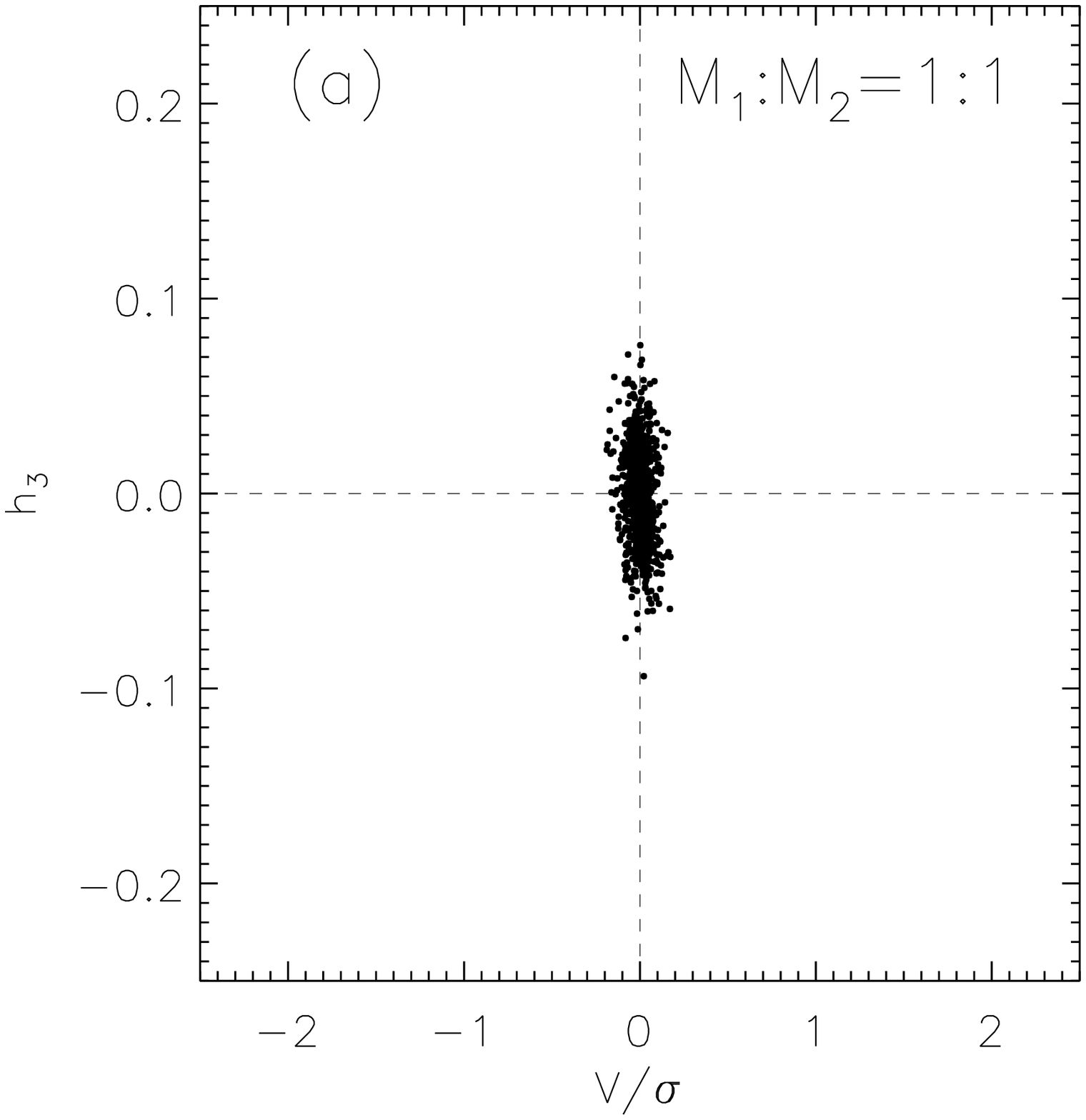}}
\hspace{0cm}
\scalebox{0.20}{ \includegraphics{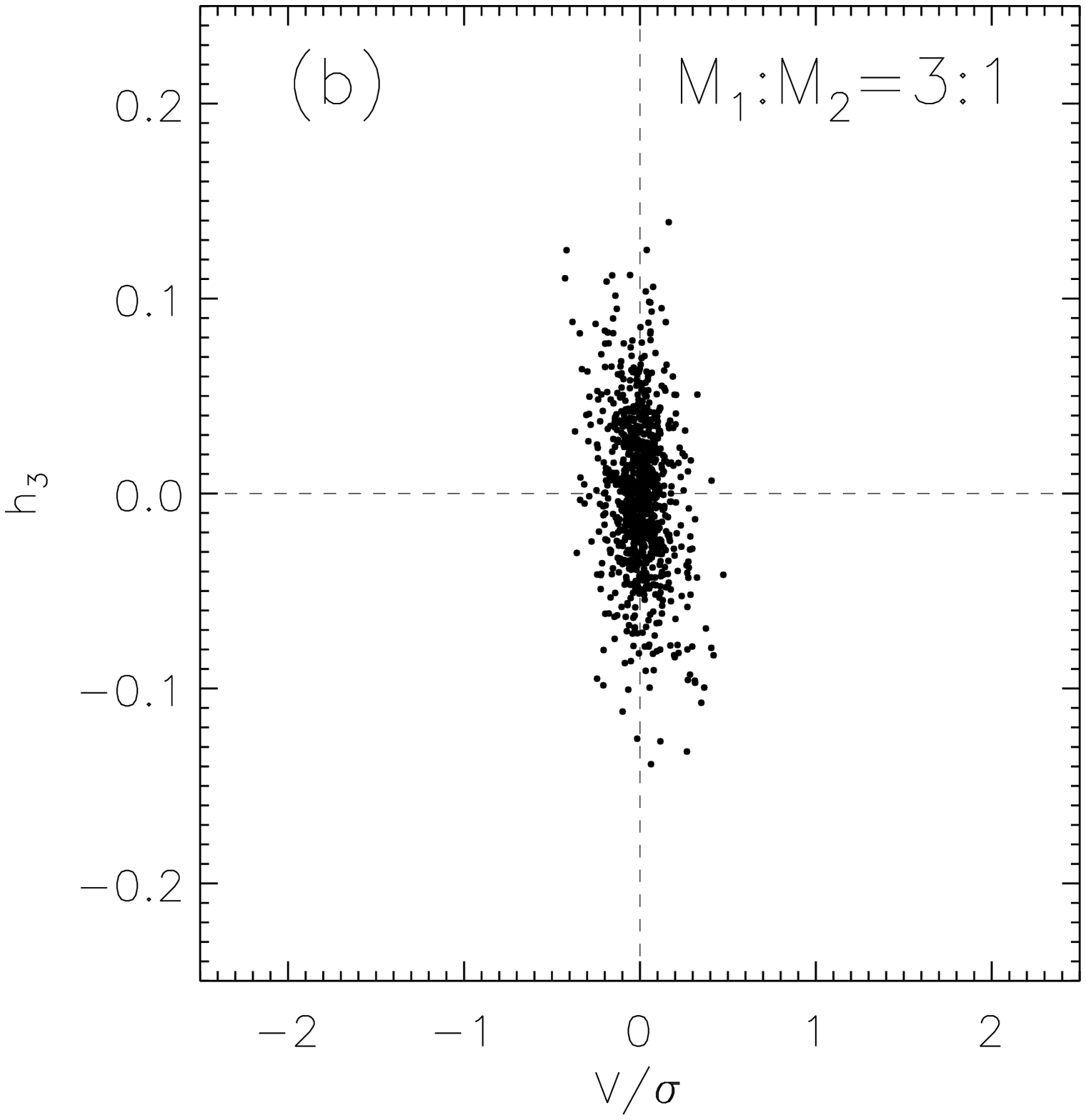}}
\vspace{0cm}
\scalebox{0.20}{ \includegraphics{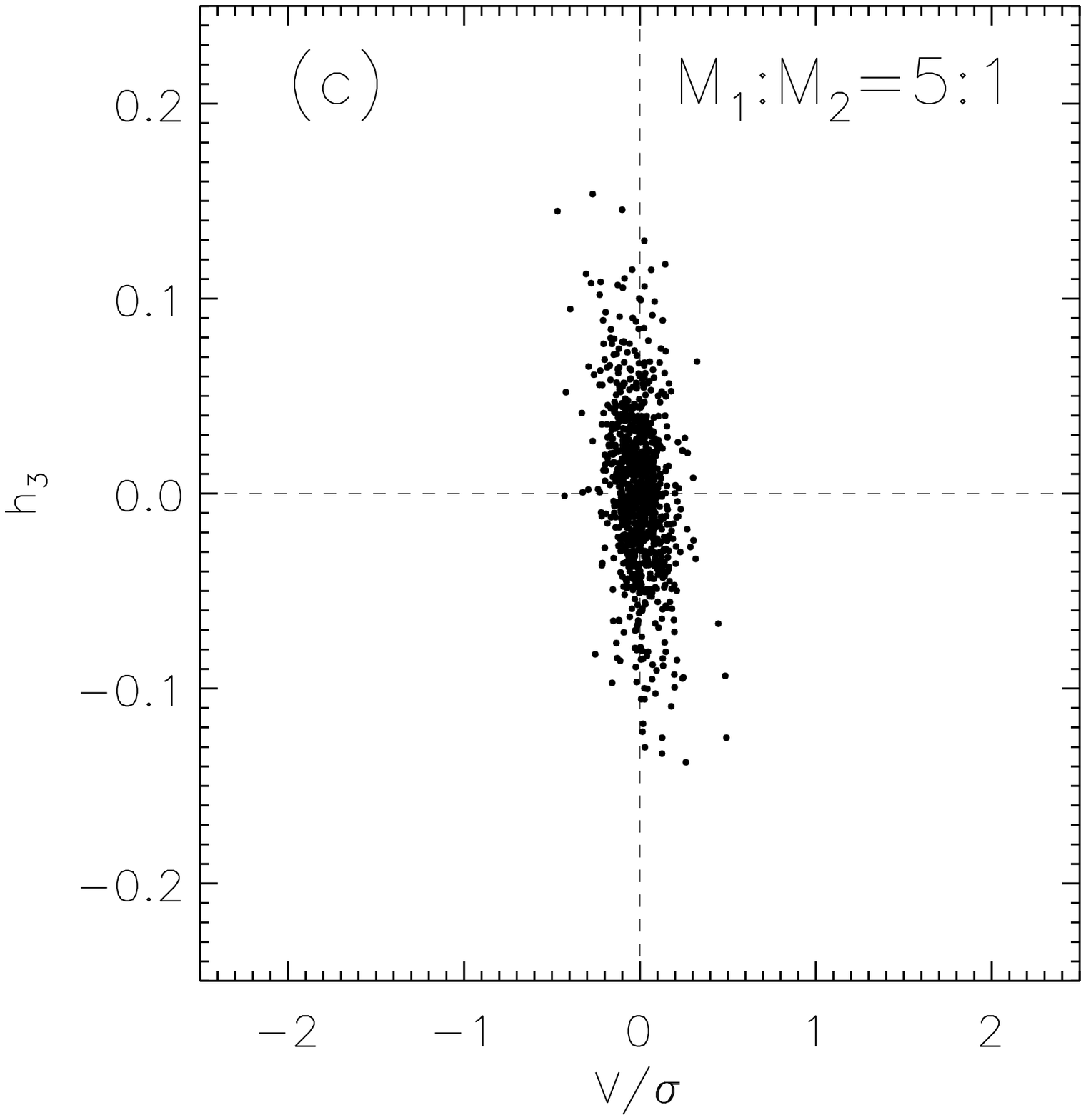}}
\hspace{0cm}
\scalebox{0.20}{ \includegraphics{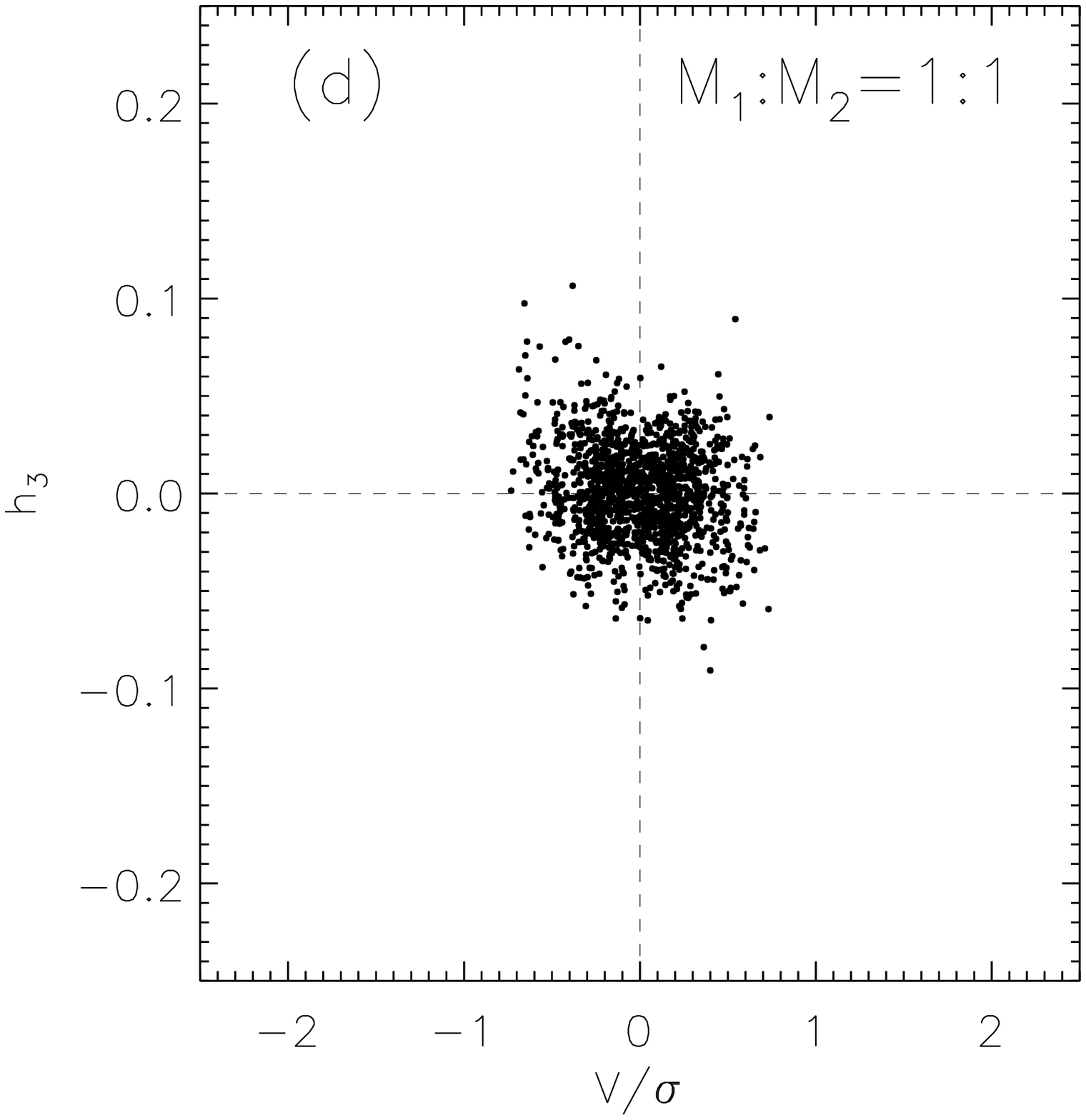}}
  \caption{$h_3$ vs. $V/\sigma$ relation for the four E+E merger models. (a) Model $e11$ with mass ratio 1:1 and small impact parameter, (b) Model $e31$ with mass ratio 3:1, (c) Model $e51$ with mass ratio 5:1 and (d)  Model $e11d$ with mass ratio 1:1 and a large impact parameter}\label{f2}
\end{figure}

\section{Results}\label{sec:results}

Following BSG, we plot together $h_3$ vs. $V/\sigma$ for all spatial bins of all slits for the merger remnants.  LOSVDs are mapped out to 1~\Reff.  
We first analyze the remnants of disc galaxy mergers, see Figure~\ref{f1}. 
Galaxy mass ratios, and bulge-to-disc ratio (B:D or D/B in the following) for the precursors, are listed in the legends.  
For equal-mass mergers (Fig.~\ref{f1}, top panels), \h3\ vs.~\Vsigma\ markedly depends on the precursors' D/B.  A small bulge (D/B=5, Fig.~\ref{f1}a) leads to low  $|\Vsigma| \leq 0.5$, and small asymmetries, $|\h3|<0.03$, with \h3\ positively correlated with \Vsigma.  Such positive correlation, opposite to that observed in real ellipticals, was also found by NB01, and would easily lead one to conclude that collisionless, equal-mass mergers of spirals yield wrong LOSVDs.  As it turns out, the LOSVDs depend on other model parameters besides the galaxy mass ratio.  
Figure~\ref{f1}b corresponds to the same equal-mass merger as in Figure~\ref{f1}a but with a larger bulge (D/B=2).  This model shows  higher rotational support within 1~\Reff, $|\Vsigma| \leq 1.3$, and higher asymmetries, $|\h3|<0.1$.  More important, \h3\ is \textit{anti}correlated with \Vsigma, except for a small region at $|\Vsigma| \sim 0$.  We conclude, in agreement with NB01, that equal-mass collisionless mergers of spirals do not reproduce the observed LOSVD asymmetries of elliptical galaxies.  But note that a significant amount of \h3\--\Vsigma\ anticorrelation can be obtained through these mergers, if the precursor host massive bulges.  

Figure~\ref{f1}c shows the results for our 3:1 merger remnant with  D/B=5.  We find larger values for \Vsigma ($\in (-2,2)$); it is well known that 3:1 S+S mergers lead to remnants with higher rotation (BB00; GGB05).   The \h3\ asymmetry term has opposite sign to \Vsigma, for high \Vsigma, whereas it has the same sign as \Vsigma\ for low \Vsigma.  The \h3 -- \Vsigma\ distribution in Figure~\ref{f1}c is quite unlike the observations.  As was the case with the equal-mass mergers, this behaviour is sensitive to D/B.  Figure~\ref{f1}d corresponds to the same model as in Figure~\ref{f1}c, except that now D/B=2.  Here the \h3\ and \Vsigma\ are anticorrelated for all values of \Vsigma. 
While the \h3 --\Vsigma\ distribution of Figure~\ref{f1}d fails to match entirely the BSG observations, we highlight that our model reproduces the correct sign for the \h3 --\Vsigma\ correlation.  Therefore, it is is not true that collisionless mergers lead to positively-correlated \h3 and \Vsigma. 

\begin{figure}
\centering
\scalebox{0.45}{ \includegraphics{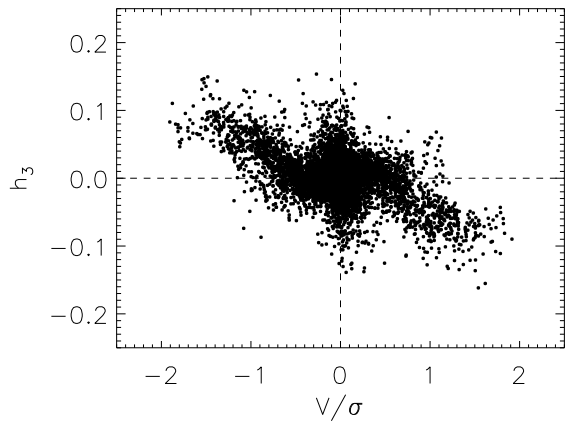}}
  \caption{Combined \h3\ vs. $V/\sigma$ relation for S+S and E+E models.}\label{f3}
\end{figure}


We argued in \S~\ref{sec:introduction} that the \h3-\Vsigma\ distribution is mostly given by the relative weights of box and short-axis tube orbits.  Matching observations (i.e., obtaining anti-correlated \h3\ and \Vsigma) requires increasing the fraction of short-axis tubes.  It is likely that our D/B=2 models reach high \h3\ and high \Vsigma\ because they have deeper central potentials than those of NB01, owing to a more massive bulge. Deeper galaxy central potentials stabilize the merging discs against bar distortions during  orbital decay.  Spin angular momentum is then deposited deeper into the remnant's luminous body, which should naturally lead to more z-tube orbits.  In contrast, discs which bar-distort during orbital decay transfer their spin angular momentum to the haloes, leading to tumbling prolate or triaxial remants with little net rotation (GGB05).  Differences in the cuspiness of the halo potentials may also lead to different relative proportions of box and z-tube orbits in ours and NB01 models.  

Two conclusions arise from the previous analysis.  First, that the \h3-\Vsigma\ distributions from S+S mergers depend on subtleties of the models such as the D/B or the shape of the inner halo potential; and, second, both ours and NB01 models suggest that the full \h3-\Vsigma\ distribution observed in elliptical galaxies (our Fig.~\ref{f0}) cannot be reproduced by S+S mergers alone.

Figure~\ref{f2} shows the results for our mergers of elliptical galaxies. The low-impact parameter mergers (Fig.~\ref{f2}abc) lead to similar distributions irrespective of the galaxy mass ratio:  low \Vsigma, large range of \h3, and \h3-\Vsigma\ anticorrelation.  These models occupy the same region of the \h3-\Vsigma\ plane as the giant, boxy, slowly-rotating ellipticals of BSG.  Model e11d (Fig.~\ref{f2}d) slightly deviates from this behaviour; \Vsigma\ reaches higher values, and \h3\ is overall smaller, and slightly uncorrelated with \Vsigma.

Thus, a way to populate the region of low $V/\sigma$ that should not be overlooked is through mergers of E's.  The outcome of these mergers show a range of properties, including diskiness (Gonzalez-Garcia \& van Albada 2005a, 2005b); model $e11d$  shows that E+E  mergers may even lead to large values of $V/\sigma$ although we must note that out elliptical galaxies models do not have a dark matter halo which would lead to a smaller transfer of angular mometum to the luminous parts (Gonzalez-Garcia \& van Albada 2005b).

What would the \h3-\Vsigma\ look like for a population of elliptical galaxies formed through mergers with varying mass ratios and types of the precursors?  
In Figure~\ref{f3} we simply overplot the \h3-\Vsigma\ distributions from all the models.  The combined distribution captures some of the most salient features of the BSG \h3-\Vsigma\ distribution of real galaxies (Fig.~\ref{f0}), namely, a cloud at low \Vsigma\ with high \h3, and two wings at high-\Vsigma, with anticorrelated \h3.  
Only one discrepancy remains, our models yield many points of view with intermediate  $V/\sigma$ and $h_3\sim 0$, which are not observed in the BSG figure.  



\section{Discussion}\label{sec:disc}

Figure~\ref{f3} suggests that the BSG \h3-\Vsigma\ distribution may be explained by S+S mergers leading to the rapidly-rotating ellipticals, and by E+E mergers leading to giant, slowly-rotating ellipticals.  There is no single collisionless origin to explain the entire BSG correlation.
But collisionless mergers, when including E+E mergers as well as S+S mergers with bulges as massive as in our models (B/D=1:2, i.e., similar to the Galaxy), explain the main features of the \h3-\Vsigma\ distribution, providing an interesting additional mechanism to that based on the NB01 assumption that a stellar disc must have formed after the merger from left-over gas.  

The \h3-\Vsigma\ relation may be seen as the imprint in the LOSVD of the angular momentum present, either of spin or of orbital origin, in the systems that merged to form present day ellipticals. Hence, LOSVDs give us information about the types of the progenitors and the orbits they were placed on. 

An interesting corollary of our explanation involving S/E precursors for the high/low \Vsigma\ domains is that giant, slowly-rotating, boxy ellipticals, which dominantly reside at low \Vsigma,  must have formed through mergers of ellipticals.  
While, at low redshifts, intermediate-mass (4:1 -- 10:1) mergers are more likely than major E+E mergers (Bournaud et al.~2005), observational evidence for the importance of E+E mergers at higher cosmological distances is mounting (van Dokkum et al. 2003, Bell et al. 2004, Tran et al. 2005, van Dokkum 2005).  Our results suggest that LOSVD asymmetries provide fossil evidence that the final steps of the merger-driven galaxy growth consisted of 'dry' mergers of ellipticals.  
Previous N-body work had already noted that low-angular momentum E+E mergers provide a clean way to reproduce many of the properties of giant, boxy ellipticals (Gonzalez-Garcia \& van Albada 2005a, 2005b; Naab, Khochfar \& Burkert 2006).  Such origin for boxy ellipticals should  be preferred over that involving bulge-less S+S mergers; the latter also lead to boxy remnants, but provide a poor match to many properties of ellipticals, including low central densities, too high axis ratios, and too high triaxiality (GGB05).  

A succession of intermediate-mass mergers, as proposed by Bournaud et al.~(2005) for the formation of ellipticals, should also be investigated as a mechanism to build remnants dominated by short-axis tubes.

Gas has been omitted from the analysis. As shown here, collisionless mergers can reproduce the most salient features of the \h3-\Vsigma\ distribution without recourse to star formation (Bekki \& Shioya 1997, NB01). 
Gas, however, must play a role in mergers of spirals, as most of the gas present in the precursors eventually returns back to the merger remnant (Hibbard \& Mihos 1995).  Given the high central concentration of merger remnants, the formation of stable, or quasi-stable, rotationally-supported gaseous discs, is not out of the question. After star formation, such discs would contribute short-axis tube orbits, as NB01 point out, thus bringing models into agreement with observations.


\section*{Acknowledgments}
We are grateful to Andreas Burkert, Thorsten Naab, Roland Jesseit and Tjeerd van Albada for useful discussions, and R.P. Saglia for kindly making his data available to us. We acknowledge the computer resources, technical expertise and assistance provided by the Barcelona Supercomputing Center, and the referee C.~J. Jog for comments that helped to improve the manuscript.

\end{document}